\documentclass[journal]{IEEEtran}
\usepackage{cite}
\usepackage{listings}
\usepackage{amsmath}
\usepackage{amsthm}
\usepackage{amssymb}
\usepackage{graphicx}
\usepackage{multirow}
\usepackage{color}
\usepackage{soul}
\usepackage{url}
\newtheorem{remark}{Remark}
\newcommand{\myfrac}{\tfrac}

\graphicspath{./figures/}

\ifCLASSINFOpdf
 
\else
 
\fi
\begin{document}

\title{Distributed Edge-to-Cloud Architecture for Continual Harmonic Load Modeling}

\author{\IEEEauthorblockN{Bhaskar Mitra and Soumya Kundu}\\
\IEEEauthorblockA{Electricity Infrastructure and Buildings Division\\ Pacific Northwest National Laboratory, Richland, WA 99354, USA\\
Email: \{bhaskar.mitra, soumya.kundu\}@pnnl.gov
}

\thanks{This work was supported by the Sensors and Data Analytics Program of the U.S. DOE Office of Electricity, under Contract No. DEAC05-76RL01830. The authors acknowledge discussions with Scott Hinson of Pecan Street Inc.}}

\maketitle

\vspace{0.0 em}
\begin{abstract}
Rapid proliferation of power electronic loads at the distribution grid edge has made accurate harmonic load modeling critical for power quality assessment, transformer derating, and grid planning. Existing Frequency Coupling Matrix (FCM) identification methods produce static, one-time models that rapidly lose fidelity as load composition evolves. This paper proposes a distributed edge-to-cloud architecture for autonomous, continual FCM re-identification under real field conditions, with two contributions: (i) a state-machine-driven pipeline that monitors point-on-wave (PoW) measurements and 
triggers FCM re-identification only when model accuracy falls below a set threshold; and (ii) an edge-to-server architecture confining waveform processing to a low-cost Raspberry Pi node, transmitting only compact JSON model updates upon confirmed change rather than streaming raw PoW data. Validated through a residential field deployment, the framework triggered retraining in 3 of 33 monitoring cycles and restored system modeling accuracy above the 80\% threshold after each event, reducing upstream data transmission by 99.7\% relative to continuous streaming.
\end{abstract}



%
\IEEEpeerreviewmaketitle

\vspace{-0.7em}
\section{Introduction}

Over the past three decades, grid-edge devices connected to distribution networks have shifted from predominantly linear to power-electronic-dominated loads, spanning electric vehicle chargers, distributed energy resources, variable frequency drives, and uninterruptible power supplies. Roughly 30\% of U.S. electric power already flowed through power electronics by 2010, and DOE projects this share could reach 80\% by 2030 \cite{ram2011power}. 
Unlike linear loads, power electronic loads draw currents with rich spectral content; their interaction with distribution network impedances produces voltage harmonics at integer multiples of the nominal frequency \cite{harmonics1983power}.

Static load representations lack the temporal resolution to capture these time-varying harmonic signatures which, if unaddressed, can drive asset degradation and uncontrolled harmonics at the feeder-head substation \cite{mcbee2013evaluating,mitra,senol2025transformer,peerzada}. 
Commercial Power Quality (PQ) meters, including low-cost smart-metering platforms, report standardized, per-channel disturbance indices such as \textit{total harmonic distortion} (THD), individual harmonic magnitudes, and IEC 61000-4-30 Class~A compliance values, but do not reveal the complex inter-dependence between voltage and current harmonics at a node \cite{smartmetering_pq2020, iec61000430}. Two principal classes of voltage-dependent harmonic load models address this gap. The Norton equivalent model represents a nonlinear load as a fixed current source in parallel with a shunt admittance \cite{mcbee2013evaluating}. The \textit{frequency coupling matrix} (FCM) \cite{Fauri_1997} is more expressive, by explicitly capturing inter-harmonic coupling between voltage and current components, a characteristic of power electronic loads \cite{grady2012understanding}. Simulation-derived FCMs \cite{lennerhag_stochastic_2020,Brunoro2017} are limited by the fidelity of their underlying device models and rarely reflect the full complexity of real deployed loads. More importantly, a one-time or batch-fit FCM only provides a snapshot but cannot track how load composition and harmonic behavior evolve over time. Addressing this requires an edge-resident framework that ingests raw point-on-wave (PoW) data, autonomously tracks load behavior, and refreshes the FCM only upon meaningful behavioral change, rather than continuously streaming PoW data upstream, which is bandwidth- and storage-prohibitive at scale. 

\textbf{Contributions:} This paper's contributions to continual harmonic load modeling are two-fold. First, an autonomous, state-machine-driven pipeline for continual FCM estimation that monitors incoming PoW measurements, detects significant shifts in load behavior, and triggers model re-identification without manual intervention. Second, a distributed edge-to-server architecture that confines raw waveform processing and FCM estimation to the edge node, transmitting only compact, validated model parameters upon confirmed behavioral change rather than streaming high-resolution waveform data. Together, these enable a scalable, bandwidth-efficient framework for autonomous harmonic load monitoring deployable across large numbers of grid-edge nodes.
%
%
In the rest, we present the FCM background and methodology (Sec.\,\ref{sec2}), the distributed edge-to-cloud architecture (Sec.\,\ref{sec3}), the experimental setup and data collection procedure (Sec.\,\ref{sec4} ), the validation results (Sec.\,\ref{sec5}), and the conclusion (Sec.\,\ref{sec6}).

\vspace{-1 em}
\section{Methodology} \label{sec2}

\subsection{FCM Estimation: Background}

The load modeling approach adopted in this paper assumes that the harmonic current 
is a linear combination of a voltage-independent base current and the frequency-coupled contributions from all voltage harmonics \cite{Fauri_1997,singhal2022harmonic}, i.e.:
\begin{align}\label{eq:FCM}
    I_h = i_{0,h} +\sum_{k=1}^{h'}Y_{h,k}V_k
\end{align}
where $I_h$ and $V_k$ are the $h$-th current and $k$-th voltage harmonic phasor, respectively; $h'$ is the highest harmonic order; $i_{0,h}$ is the $h$-th voltage-independent base current harmonic phasor, and $Y_{h,k}$ are cross-harmonic admittances linking $k$-th voltage harmonic to $h$-th current harmonic. FCM estimation requires identification of the base current components $i_{0,h}$ and the cross-harmonic admittances $Y_{h,k}$\,.


Several pairs of voltage and current measurements from the PoW sensor are stacked column-wise to yield:
%
\begin{equation}\label{eqn2:FCMexp}
\begin{split}
    \underbrace{\begin{bmatrix} I_h^{(1)} \\ I_h^{(2)} \\ \vdots \\ I_h^{(M)} \end{bmatrix}}_{\mathbf{I}_h \,\in\, \mathbb{C}^M}
    &=
    \underbrace{\begin{bmatrix} 
        1 & V_1^{(1)} & V_2^{(1)} & \dots  & V_{h'}^{(1)} \\ 
        1 & V_1^{(2)} & V_2^{(2)} & \dots  & V_{h'}^{(2)} \\ 
        \vdots & \vdots & \vdots & \ddots & \vdots \\ 
        1 & V_1^{(M)} & V_2^{(M)} & \dots  & V_{h'}^{(M)} 
    \end{bmatrix}}_{\mathbf{Z} \,\in\, \mathbb{C}^{M \times (h'+1)}}
    \underbrace{\begin{bmatrix} 
        i_{0,h} \\ Y_{h,1} \\ Y_{h,2} \\ \vdots \\ Y_{h,h'} 
    \end{bmatrix}}_{\mathbf{Y}_h \,\in\, \mathbb{C}^{h'+1}} 
\end{split}
\end{equation}
where,
$M$ is the number of measurement snapshots;
    $\mathbf{I}_h \in \mathbb{C}^M $ is the column vector of measured harmonic current phasor at order $h$ across all $M$ snapshots; $\mathbf{Z} \in \mathbb{C}^{M \times (h'+1)}$ is the measurement matrix whose $m$-th row contains the voltage-independent unit entry followed by the $h'$ complex voltage harmonic phasors at snapshot $m$; $\mathbf{Y}_h \in \mathbb{C}^{h'+1}$ is the FCM parameter column vector containing the base current source $i_{0,h}$ and the $h'$ admittance coupling coefficients $Y_{h,1}, \dots, Y_{h,h'}$
 
\begin{remark}
    The equation \eqref{eqn2:FCMexp} describes a single phase system, but can be extended to three-phase installations by constructing and solving an independent instance of \eqref{eqn2:FCMexp} for each phase.
    Such per-phase decoupling is valid under negligible inter-phase harmonic coupling, typical of standard distribution system operating conditions with mild voltage imbalance.
\end{remark} 

To satisfy the identifiability condition, the number of discrete measurements $M \ge (h'+1)$ and 
the system must be overdetermined, $M > (h'+1)$, with sufficient linear independence across the rows of $\mathbf{Z}$; estimation robustness improves as $M$ increases. In this study, $h'=10$ and $M=15$, a modest margin over the 11 unknowns per harmonic; identifiability therefore relies on the snapshots spanning distinct operating windows, and the held-out validation accuracy in Sec.\,\ref{sec5} confirms that the resulting estimate generalizes.
With an overdetermined system, we can estimate the optimal coefficients of $\widehat{\mathbf{Y}}_h$ (for each harmonic order $h$) by minimizing the sum of the squared residuals:
\begin{equation}
\widehat{\mathbf{Y}}_h = \arg \min_{\mathbf{Y}_h} \left\|\mathbf{Z} \mathbf{Y}_h - \mathbf{I}_h\right\|_2^2= (\mathbf{Z}^\dagger\mathbf{Z})^{-1}\mathbf{Z}^\dagger\mathbf{I}_h\label{eqn:least}
\end{equation}
where the closed-form solution uses pseudoinverse of $\mathbf{Z}_h$, and $^\dagger$ denotes the adjoint. The FCM up to order $h'$ is given by:
\begin{align}
    \texttt{FCM}_{h'} = \begin{bmatrix}
        \widehat{\mathbf{Y}}_1 & \widehat{\mathbf{Y}}_2 & \widehat{\mathbf{Y}}_3 &\dots & \widehat{\mathbf{Y}}_{h'}
    \end{bmatrix}
\end{align}

In \cite{singhal2022harmonic} authors highlight the need for a quantifiable set of measurements to estimate the FCM. The proposed framework directly overcomes this limitation by deploying an edge-resident sensor that captures continuous, high-resolution raw power waveforms at the point of load, providing direct access to the full harmonic spectrum.
 

\vspace{-0.5em}
\subsection{Reconstruction Error Calculation}
To evaluate the efficacy and precision of the proposed architecture, validation metrics were established based on both harmonic current magnitudes and active power predictions. The $h$-th harmonic order of the current vector ($\widehat{I}_h$) is computed from the FCM parameters, and the measured voltage harmonics, following equation \eqref{eq:FCM}: 
\begin{align}\label{eq:prediction}
    \widehat{I}_h=\begin{bmatrix}
        1 & V_1 & V_2 &\dots &V_{h'}
    \end{bmatrix}\,\widehat{\mathbf{Y}}_h
\end{align}
Both the measured and predicted harmonic current vectors are converted from raw discrete Fourier transform bin values to physical amplitude units by normalizing against the number of time-domain samples in the captured window, $N$.





The system modeling accuracy($\mathcal{A}_I$) is quantified using the relative $L_2$-norm error, computed as the $L_2$-norm of the current estimation errors across all $h'$ considered harmonics, normalized against the $L_2$-norm of the measured harmonic current vector:
%
\begin{equation}
    \mathcal{A}_I := \left(1-\sqrt{\myfrac{\sum_{h=1}^{h'}\left|I_h\!-\!\widehat{I}_h\right|^2}{\sum_{h=1}^{h'}\left|I_h\right|^2}}  \right) \label{accuracy}
\end{equation}


Additionally, the harmonic active power is evaluated to ascertain the modeling accuracy in capturing phase relationships, a critical requirement for power quality assessment. The active power for each harmonic $h$ is computed for both measured ($P_h$) and predicted ($\widehat{P}_h$) states using the following,
\begin{equation}
    P_h = \texttt{real}(V_h I_h^*)\,,\,~\,\widehat{P}_h = \texttt{real}(V_h \widehat{I}_h^*) \label{power}
\end{equation}
where $^*$ (super-script) denotes the complex conjugate of the phasor. The per-harmonic active power error ($\mathcal{E}_{P,h}$) is normalized against the measured fundamental active power ($P_1$):
\begin{align}
    \mathcal{E}_{P,h} = \myfrac{|P_h\!-\!\widehat{P}_h|}{|P_1|}
\end{align}
while the overall active power accuracy ($\mathcal{A}_P$) is defined as:
\begin{align}
    \mathcal{A}_P := \left(1-\sqrt{\myfrac{\sum_{h=1}^{h'}\left|P_h\!-\!\widehat{P}_h\right|^2}{\sum_{h=1}^{h'}\left|P_h\right|^2}}  \right) \label{accuracy_power}
\end{align}
Because (\ref{accuracy_power}) normalizes by the harmonic active powers, which are small relative to $P_1$, $A_P$ is a stricter metric than $\mathcal{E}_{P,h}$.

\begin{figure*}[t]
    \centering
    \includegraphics[width=\linewidth]{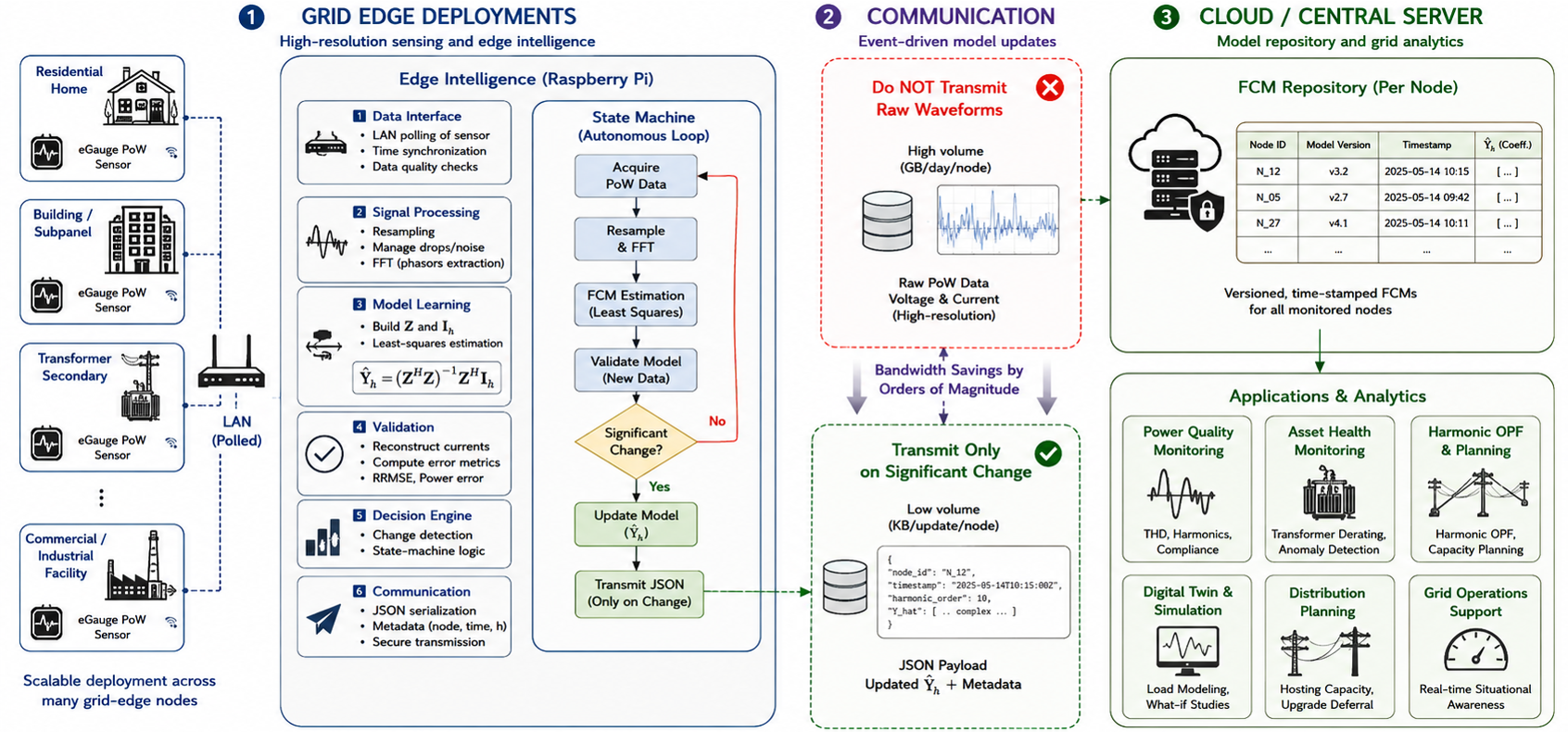}
    \caption{Proposed edge-to-cloud architecture for continual Frequency Coupling Matrix (FCM) estimation. Raw point-on-wave (PoW) measurements are processed locally at the grid edge to estimate and validate FCMs. Only validated model updates are transmitted to the central server upon significant behavioral changes, enabling scalable, bandwidth-efficient harmonic load modeling and downstream grid analytics.}
    \label{fig:concept}
\end{figure*}

\vspace{-0.5em}
\section{Proposed Continual Modeling Architecture} \label{sec3}

The proposed continual load modeling framework is realized through a three-tier distributed architecture (see Fig. \ref{fig:concept}) that separates the computationally intensive tasks of PoW data acquisition and model identification from the lightweight responsibilities of model aggregation and storage. Processing at the edge and transmitting only validated models, rather than raw waveforms, is central to achieving scalability across large numbers of monitoring nodes without imposing prohibitive demands on network infrastructure.

\vspace{-0.5em}

\subsection{Hardware Setup}

The physical instrumentation layer is built around a PoW sensor that serves as the primary data acquisition front-end at each monitored grid-edge node. In this study, commercially available, revenue-grade power meters, eGauge sensors 
(EG4015~\cite{egauge_core,egauge_split_core_ctid_online})
are used to capture high-resolution voltage and current waveforms at a sampling resolution of 2.48~kHz. The sensors are interfaced to a local edge compute node implemented on a Raspberry Pi 
(Model 4B~\cite{raspberry_pi_4_online})
over a local area network (LAN) connection, enabling programmatic polling of raw waveform records at configurable intervals. 
The Raspberry Pi was selected for its low cost, modest power draw, and 
sufficient throughput for the workloads in Sec.~\ref{sec:edge node}, supporting scalable 
deployment across distribution assets (e.g., transformers, sub-panels, building entry points).

\vspace{-1em}
\subsection{Edge Node} \label{sec:edge node}


The edge compute node constitutes the computational core of the proposed architecture, executing the full FCM identification pipeline locally and autonomously without dependence on upstream connectivity during normal operation. Upon polling a new batch of raw voltage and current waveform records from the eGauge meter, the edge node first resamples the data to factor out data drops, noise, and bad samples. Upon successful resampling, the node uses \textit{Fast Fourier Transform} (FFT) to decompose the time-domain signals into their constituent harmonic phasors, extracting the complex-valued voltage ($V^{(m)}_h$) and current ($I^{(m)}_h$) components at each harmonic order-$h$ for measurement snapshot $m$. These spectral estimates populate the measurement matrix ($\mathbf{Z}$) and the current observation vector $\mathbf{I}_h$, as discussed in Sec.\,\ref{sec2}.

On accumulating sufficient PoW data, the FCM coefficients are resolved via least squares \eqref{eqn:least}, and the resulting model is evaluated against a held-out set of PoW data to validate its efficacy. The continual framework operates by polling the sensor at a pre-determined interval and re-evaluating system modeling accuracy against each new batch, with the retraining and transmission criteria formalized in Sec.\,\ref{sec:change_detection}.

\vspace{-1.2em}
\subsection{Change Detection and Retraining} \label{sec:change_detection}

The continual estimation loop is governed by a threshold-based change detection criterion applied to the system modeling accuracy defined in (\ref{accuracy}). Following each FCM identification cycle, the estimated model parameters ($\widehat{\mathbf{Y}}_h$) are evaluated 
against a held-out validation set of PoW measurements not used during training (a separate 4\,s window per validation run). If the resulting $\mathcal{A}_I$ falls below
a pre-determined threshold $\tau_{acc}$, the current model is deemed no longer representative of the prevailing load behavior and a retraining event is triggered:
\begin{equation}
    \texttt{retrain FCM if}\,~\,\mathcal{A}_I < \tau_{acc}
\end{equation}
where $\mathcal{A}_I$ is computed per \eqref{accuracy} on the held-out validation set. Upon triggering, the edge node collects a fresh set of $M \geq (h'+1)$ linearly independent measurement snapshots, re-estimates $\widehat{\mathbf{Y}}_h$ via (\ref{eqn:least}), re-validates against a new held-out set, and serializes the updated coefficients for transmission to the central server only if the re-validated $\mathcal{A}_I$ meets or exceeds $\tau_{acc}$. If the criterion is satisfied,
the edge node continues polling without transmitting, preserving bandwidth. The threshold $\tau_{acc}$ is a configurable parameter. A higher threshold increases sensitivity but causes retraining on minor, transient load changes and a lower threshold suppresses such events but delays the response to genuine changes in load composition. The 80\% value balances these effects for the residential site studied and should be tuned to the target load class.

\vspace{-0.7 em}
\subsection{Central Server}
The central server operates as a lightweight model aggregation layer. Continuous streaming of high-resolution waveform data would generate data volumes on the order of gigabytes (GB) per node per day; instead, the edge node transmits compact JSON payloads containing only the estimated FCM coefficient vectors $\widehat{\mathbf{Y}}_h$ and associated metadata, sent exclusively upon detection of a significant load behavioral change. Each payload encodes the full set of admittance coupling coefficients across the monitored harmonic orders, the harmonic index $h$, a timestamp, and a node identifier, constituting a complete and self-contained model update that can be directly ingested into the server-side model registry without further processing. 
This event-driven protocol dramatically reduces upstream bandwidth consumption (quantified in Sec. V) and keeps the server-side infrastructure agnostic to the number of edge nodes, enabling scaling across large distribution service territories. The aggregated FCM repository maintained at the server provides a longitudinal record of load evolution across the monitored network, supporting downstream applications in power quality assessment, transformer derating analysis, and related grid planning tasks.

In this study, the central server role is fulfilled by a local workstation on 
the same LAN, serving as a proof-of-concept receiver; the JSON payload format and transmission protocol are otherwise hardware-agnostic and equally 
applicable to cloud, SCADA, or any HTTP/MQTT endpoint.
\vspace{-0.7em}
\section{Experimental Setup} \label{sec4}

Validation was performed through a field deployment at a single-family residential home served by a standard split-phase, 240 V line-to-line (L-L) service.
Data acquisition was done using an eGauge Core energy meter \cite{egauge_core}, a 15-channel multi-circuit smart meter natively supporting up to three-phase service and logging voltage, current, power, frequency, and harmonic distortion at sub-second resolution. Current sensing was provided by two 100\,A split-core current transformers (CTs) with CTid auto-detection\cite{egauge_split_core_ctid_online}, 
clamped around the two incoming line conductors feeding the residential panel. The meter logs at its finest available granularity for the recording window used in this study, providing high-resolution waveform data required for subsequent FFT-based harmonic decomposition.
The eGauge meter connects to the Raspberry Pi edge node over a local area network (LAN), exposing measurement data through a native JSON API that the edge node polls directly without needing proprietary drivers. 

Voltage and current waveforms were acquired from the eGauge Core meter at a native sampling rate of approximately 2.48\,kHz. Prior to harmonic analysis, the raw waveforms were resampled to a uniform rate of 2.4 kHz which is an integer multiple of the 60 Hz. 
%
%
Each resampled waveform window was then processed through FFT to extract the complex-valued voltage ($V_h^{(m)}$) and current ($I_h^{(m)}$) phasors at each harmonic order. Harmonic contents up to (including) $10^{th}$ order were retained for FCM identification.

\vspace{-0.5em}
\section{Results} \label{sec5}

This section presents experimental results from the field deployment described in Sec.\,\ref{sec4}, evaluating the proposed continual FCM estimation framework along two dimensions: system modeling accuracy and system operational behavior.

\begin{figure}
    \centering
    \includegraphics[width=\linewidth]{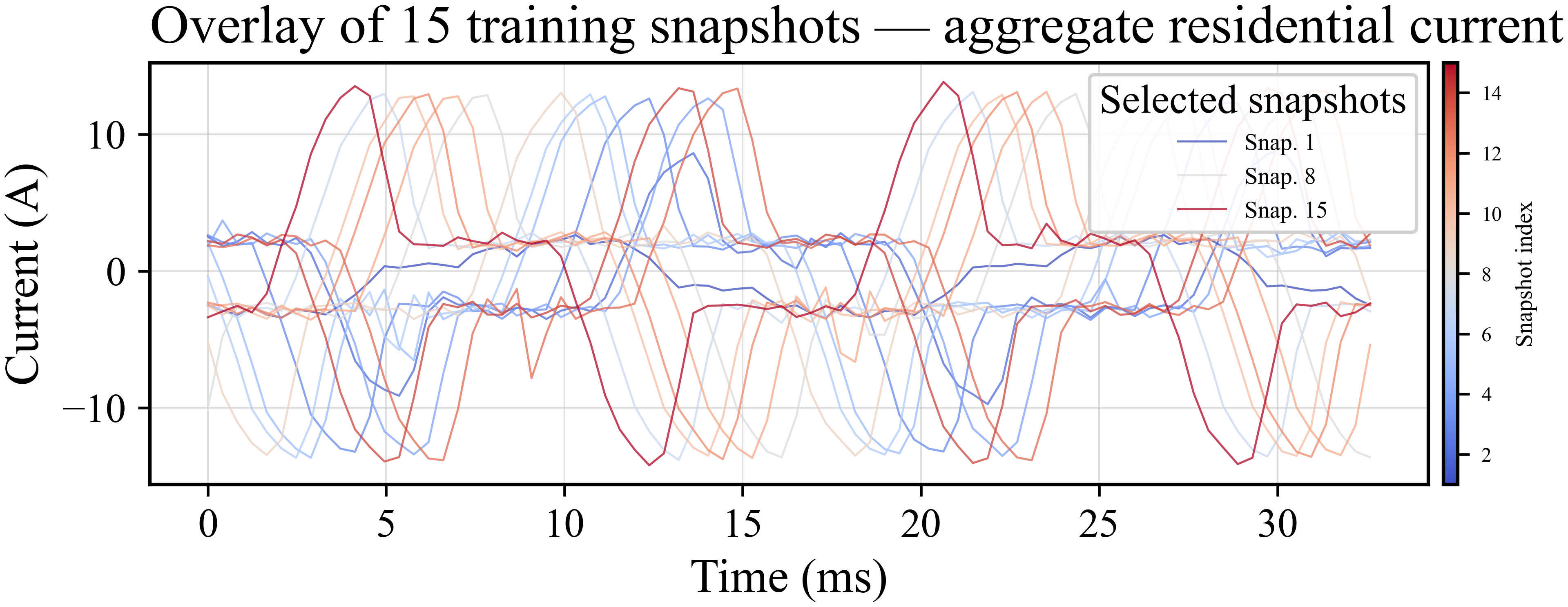}
    \caption{Overlay of all 15 training current snapshots captured at the residential service entrance, each representing a distinct steady-state operating window.}
    \label{fig:snapshot}
\end{figure}

\begin{figure}
    \centering
    \includegraphics[width=0.8\linewidth]{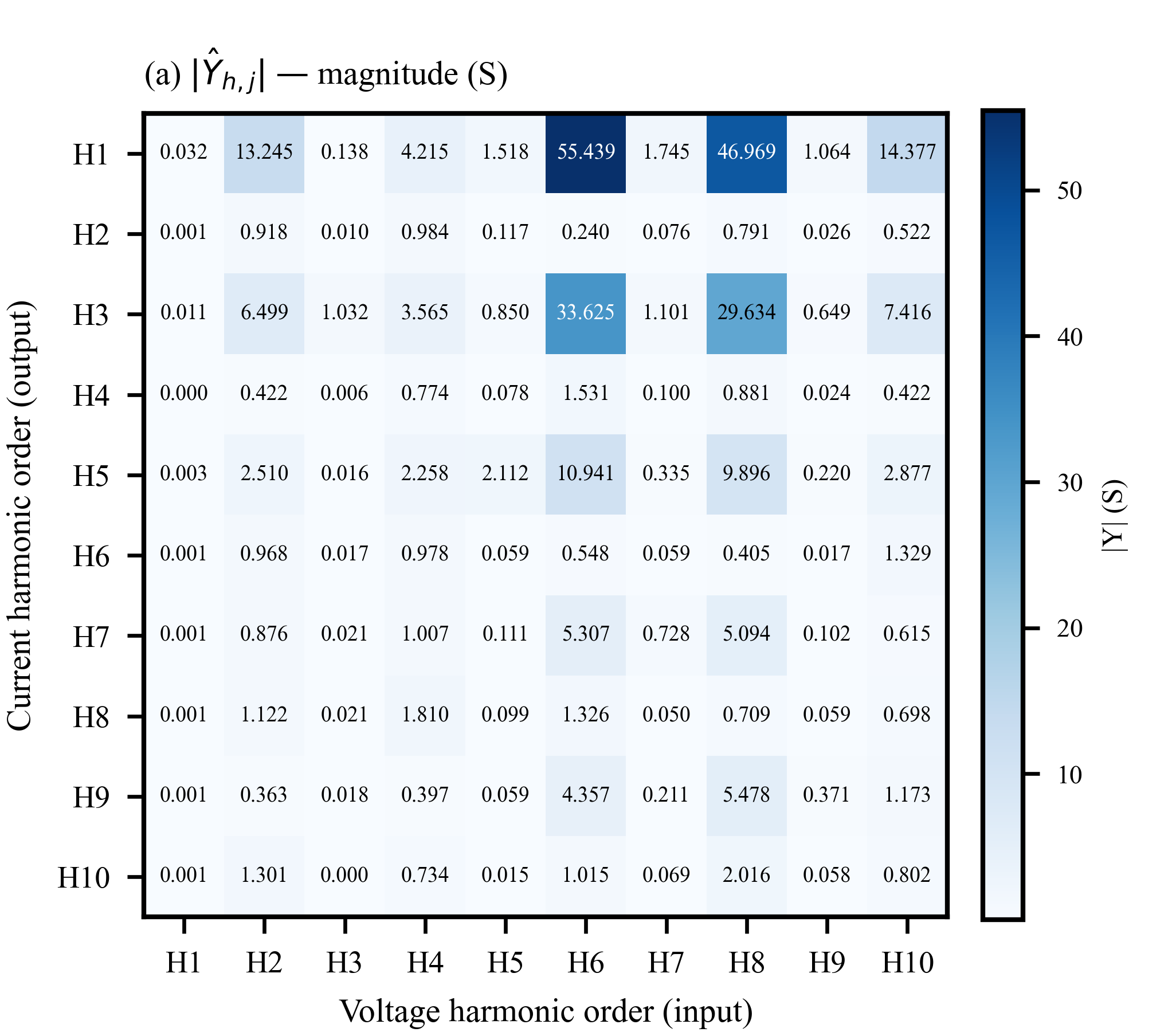}
    \caption{Estimated FCM admittance matrix ($\widehat{\mathbf{Y}}_h$) absolute values (element-wise) for the residential test site: with harmonic output (current) order on the vertical axis and harmonic input (voltage) order on the horizontal axis. Non-negligible off-diagonal entries confirm cross-harmonic coupling.}
    \label{fig:heatmap}
\end{figure}

Fifteen independent voltage and current measurement snapshots, each spanning three cycles, were collected to estimate $\widehat{\mathbf{Y}}_h$ via (\ref{eqn:least}), shown in Fig. \ref{fig:snapshot}. Fig. \ref{fig:heatmap} presents a heatmap of the resulting FCM admittance matrix magnitude (element-wise absolute values), revealing 
several non-trivial off-diagonal entries at lower harmonic orders
that indicate measurable cross-harmonic coupling between voltage and current at the edge node, 
consistent with the switching behavior of residential power electronic loads. Presence of these cross-harmonic coupling terms underscores the necessity of the FCM framework over alternatives such as a Norton equivalent model.

\begin{figure}
    \centering
    \includegraphics[width=\linewidth]{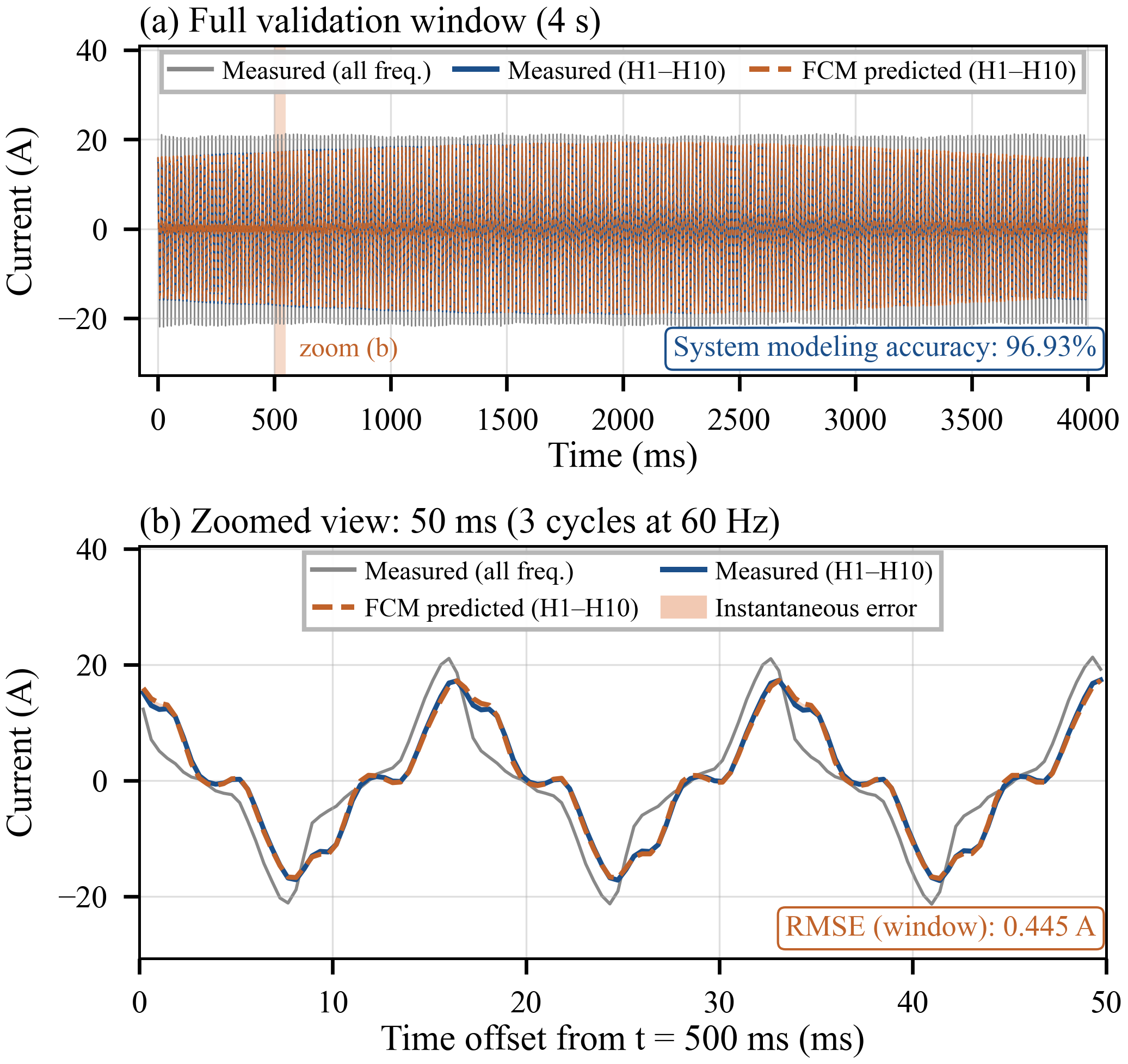}
    \caption{FCM validation results for the residential test site: (a) 4-sec test window showing raw measured current (gray), harmonic-filtered measured current (blue, up to 10-th order), and FCM-reconstructed predicted current (orange dashed), with the magnified region highlighted; (b) zoomed view of a representative 50~ms window.}
    \label{fig:zoom}
\end{figure}

\begin{figure}
    \centering
    \includegraphics[width=\linewidth]{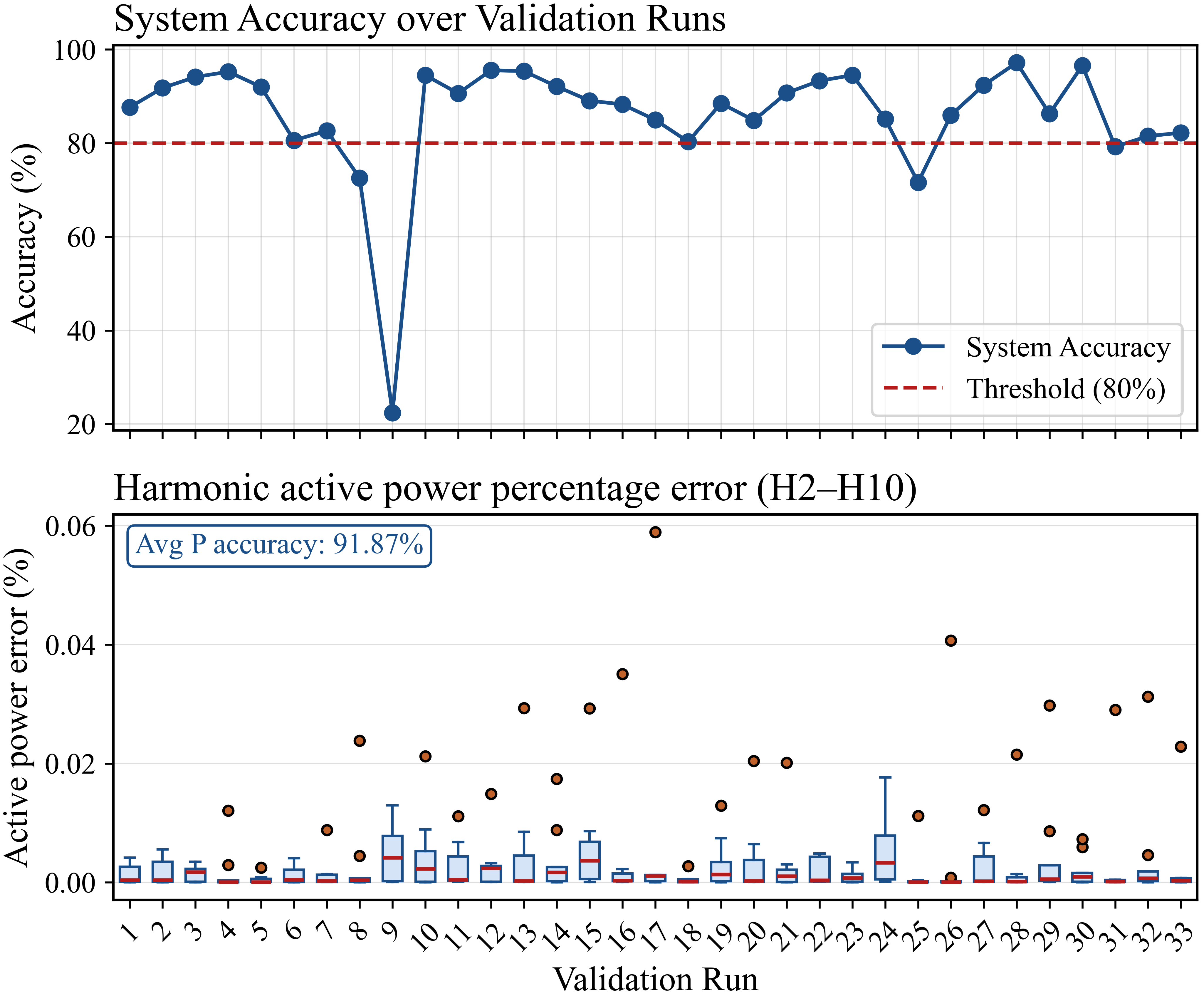}
    \caption{Continual FCM estimation performance over 33 consecutive validation runs: (upper) system modeling accuracy $\mathcal{A}_I$ relative to the retraining threshold ($\tau_{acc}$) of 80\%, with 3 retraining events visible as accuracy recoveries following threshold violations; (lower) per-harmonic, normalized, active power error ($\mathcal{E}_{P,h}$) remaining below 1\% across all validation runs.}
    \label{fig:error}
\end{figure}

Fig. \ref{fig:zoom} shows validation against an independent measurement set, demonstrating successful reconstruction of the current waveform over a separate 4\,s data. Fig. \ref{fig:error} shows continual model performance across 33 consecutive validation runs, tracking system accuracy ($\mathcal{A}_I$) against the retraining threshold (at 80\,\%) alongside per-harmonic active power percentage error ($\mathcal{E}_{P,h}$, up to $10^{th}$ order). 
System accuracy (upper subplot) stays above $\tau_{acc}$ for most runs;
threshold violations trigger the retraining procedure of
Sec.\,\ref{sec:change_detection} using a fresh set of 15 snapshots,
with 3 retraining events visible as accuracy recoveries above $\tau_{acc}$.
The harmonic active power percentage error (lower subplot) remains below 1\% across all validation runs, normalized against the measured fundamental active power $P_1$, confirming that the estimated FCM captures both harmonic current magnitude and the phase relationships needed for reliable active power prediction under distorted grid conditions.

A principal contribution of the proposed architecture is the reduction in upstream data transmission achieved by the event-driven, model-centric protocol relative to continuous raw PoW streaming. Under continuous streaming, each validation run requires transmission of the full raw PoW waveform, approximately 150~KB per run, or 4.95~MB across the full 33-run monitoring window. Under the proposed protocol, the edge node transmits a JSON payload of approximately 5~KB only upon a confirmed behavioral change. Since only 3 of the 33 runs triggered retraining, total upstream data volume was approximately 15~KB, a 99.7\% reduction relative to continuous streaming. Table \ref{tab:compare} summarizes this comparison.

\vspace{-0.5em}

\vspace{0em}

\begin{table}[!t]
\caption{Upstream data transmission over 33 validation runs}
\label{tab:compare}
\begin{tabular}{|l|c|c|}
\hline
\multicolumn{1}{|c|}{\multirow{2}{*}{\textbf{Metric}}} & \textbf{Continuous PoW} & \textbf{Proposed}     \\ \cline{2-3} 
\multicolumn{1}{|c|}{}                                 & \textbf{Streaming}      & \textbf{Event-Driven} \\ \hline
Raw data per validation run                            & $\sim$150 KB            & 0 KB                  \\ \hline
JSON payload per transmission                          & —                       & $\sim$5 KB            \\ \hline
Transmission events (33 runs)                          & 33                      & 3                     \\ \hline
Total data transmitted                                 & $\sim$4.95 MB           & $\sim$15 KB           \\ \hline
\end{tabular}
\end{table}

\vspace{-0.5em}

\section{Conclusion} \label{sec6}
This paper presented an autonomous, edge-deployed framework for continual FCM estimation of grid-edge harmonic loads. In a residential field deployment over 33 monitoring runs, the Raspberry Pi edge node detected three significant changes in load behavior, retrained autonomously, and restored system modeling accuracy above the 80\% threshold after each event. It reconstructed harmonic currents up to the 10th order with per-harmonic active power error below 1\% of the fundamental and captured cross-harmonic coupling that Norton equivalent models cannot represent. Transmitting only validated JSON model updates reduced upstream data from $\sim$4.95\,MB to $\sim$15\,KB (99.7\%) relative to continuous PoW streaming. Future work will extend validation to multiple sites and load classes, evaluate fleet-scale communication, and study adaptive thresholds.
%


%

\ifCLASSOPTIONcaptionsoff
  \newpage
\fi



\bibliographystyle{IEEEtran}
\bibliography{IEEEabrv,./bib/paper}

\end{document}